\documentclass[12pt]{article}
\usepackage{graphicx}

\def\Re{{\cal R \mskip-4mu \lower.1ex \hbox{\it e}\,}}
\def\Im{{\cal I \mskip-5mu \lower.1ex \hbox{\it m}\,}}
\def\ie{{\it i.e.}}
\def\eg{{\it e.g.}}

\def\sub#1{_{\lower.25ex\hbox{$\scriptstyle#1$}}}
\def\tev{\,{\ifmmode\mathrm {TeV}\else TeV\fi}}
\def\gev{\,{\ifmmode\mathrm {GeV}\else GeV\fi}}
\def\mev{\,{\ifmmode\mathrm {MeV}\else MeV\fi}}
\def\mpl{\ifmmode M_{pl}\else $M_{pl}$\fi}
\def\mpl{\ifmmode \overline M_{Pl}\else $\bar M_{Pl}$\fi}
\def\to{\rightarrow}

\def\subw{_{\rm w}}
\def\mh{\ifmmode m\sbl H \else $m\sbl H$\fi}
\def\mch{\ifmmode m_{H^\pm} \else $m_{H^\pm}$\fi}
\def\mt{\ifmmode m_t\else $m_t$\fi}
\def\mc{\ifmmode m_c\else $m_c$\fi}
\def\mz{\ifmmode M_Z\else $M_Z$\fi}
\def\mw{\ifmmode M_W\else $M_W$\fi}
\def\mws{\ifmmode M_W^2 \else $M_W^2$\fi}
\def\mhs{\ifmmode m_H^2 \else $m_H^2$\fi}   
\def\mzs{\ifmmode M_Z^2 \else $M_Z^2$\fi}
\def\mts{\ifmmode m_t^2 \else $m_t^2$\fi}
\def\mcs{\ifmmode m_c^2 \else $m_c^2$\fi}
\def\mchs{\ifmmode m_{H^\pm}^2 \else $m_{H^\pm}^2$\fi}
\def\ztwo{\ifmmode Z_2\else $Z_2$\fi}
\def\zone{\ifmmode Z_1\else $Z_1$\fi}
\def\mtwo{\ifmmode M_2\else $M_2$\fi}
\def\mone{\ifmmode M_1\else $M_1$\fi}
\def\tb{\ifmmode \tan\beta \else $\tan\beta$\fi}
\def\xw{\ifmmode x\subw\else $x\subw$\fi}
\def\ch{\ifmmode H^\pm \else $H^\pm$\fi}
\def\lum{\ifmmode {\cal L}\else ${\cal L}$\fi}
\def\inpb{\,{\ifmmode {\mathrm {pb}}^{-1}\else ${\mathrm {pb}}^{-1}$\fi}}
\def\infb{\,{\ifmmode {\mathrm {fb}}^{-1}\else ${\mathrm {fb}}^{-1}$\fi}}
\def\epem{\ifmmode e^+e^-\else $e^+e^-$\fi}
\def\ppb{\ifmmode \bar pp\else $\bar pp$\fi}
\def\bsg{\ifmmode B\to X_s\gamma\else $B\to X_s\gamma$\fi}
\def\bsll{\ifmmode B\to X_s\ell^+\ell^-\else $B\to X_s\ell^+\ell^-$\fi}
\def\bstt{\ifmmode B\to X_s\tau^+\tau^-\else $B\to X_s\tau^+\tau^-$\fi}
\def\lamt{\ifmmode \tilde\lambda\else $\tilde\lambda$\fi}
\def\shat{\ifmmode \hat s\else $\hat s$\fi}
\def\that{\ifmmode \hat t\else $\hat t$\fi}
\def\uhat{\ifmmode \hat u\else $\hat u$\fi}

\newskip\zatskip \zatskip=0pt plus0pt minus0pt
\def\matth{\mathsurround=0pt}
\def\lsim{\mathrel{\mathpalette\atversim<}}

\def\atversim#1#2{\lower0.7ex\vbox{\baselineskip\zatskip\lineskip\zatskip
  \lineskiplimit 0pt\ialign{$\matth#1\hfil##\hfil$\crcr#2\crcr\sim\crcr}}}

\def\grtsim{\,\,\rlap{\raise 3pt\hbox{$>$}}{\lower 3pt\hbox{$\sim$}}\,\,}
\def\lsim{\,\,\rlap{\raise 3pt\hbox{$<$}}{\lower 3pt\hbox{$\sim$}}\,\,}

\renewcommand{\thefootnote}{\fnsymbol{footnote}}

\begin{document} \begin{titlepage}
\rightline{\vbox{\halign{&#\hfil\cr
&SLAC-PUB-14954\cr
}}}
\begin{center}
\thispagestyle{empty} \flushbottom { {\Large\bf Possible Suppression of Resonant Signals for Split-UED by Mixing at the LHC? 
\footnote{Work supported in part by the Department of Energy, Contract DE-AC02-76SF00515}
\footnote{e-mail:rizzo@slac.stanford.edu}}}
\medskip
\end{center}

\centerline{Thomas G. Rizzo}
\vspace{8pt} 
\centerline{\it SLAC National
Accelerator Laboratory, 2575 Sand Hill Rd., Menlo Park, CA, 94025}

\vspace*{0.3cm}

\begin{abstract}
The mixing of the imaginary parts of the transition amplitudes of nearby resonances via the breakdown of the Breit-Wigner approximation has been shown 
to lead to potentially large modifications in the signal rates for new physics at colliders. In the case of suppression, this effect may be significant enough to 
lead to some new physics signatures being initially missed in searches at, \eg, the LHC. Here we explore the influence of this `width mixing' on the production of the nearly degenerate, 
level-2 Kaluza-Klein (KK) neutral gauge bosons present in Split-UED. We demonstrate that in this particular case large cross section modifications in 
the resonance region are necessarily absent and explain why this is so based on the group theoretical structure of the SM.   
\end{abstract}



\renewcommand{\thefootnote}{\arabic{footnote}} \end{titlepage} 

%
%
%
%
%

\section{Introduction and Background}

Models of new physics occasionally allow for the existence of approximately degenerate states which under reasonable assumptions are likely to share 
common decay modes leading to unanticipated effects. Some potential examples of such possibilities include, \eg, $t$ and $t'$ quarks in fourth generation models{\cite{Pilaftsis:1997dr}}, 
the heavy Higgs fields,  $H$ and $A$, in the CP-violating MSSM{\cite{Papavassiliou:1997pb}}{\cite{Ellis:2005fp}}, CP-violating effects in neutrino mixing between nearly degenerate 
states{\cite{Bray:2007ru}}, the lightest $Z'$ and $A'$ Kaluza-Klein (KK) fields in Higgsless models{\cite{Cacciapaglia:2009ic}} as well as the level-2 KK neutral gauge bosons, 
$\sim W^0_2$ and $\sim B^0_2$, present in Split-Universal Extra Dimensions (\ie, Split-UED). Frequently, as in the later two examples, these are states that can be searched for 
as resonances in a particular channel, such as in the Drell-Yan process, \ie, $pp\to \ell^+\ell^- +X$, at the LHC. In such searches the mutual effects of these dual 
resonances upon each other can play an important role. 

In pioneering work, the authors of Refs.{\cite{Nowakowski:1993iu}} and {\cite{Cacciapaglia:2009ic}} have demonstrated that the necessary conditions for a breakdown 
in the usual double Breit-Wigner (BW) description of such nearly degenerate resonance pairs to occur are:  ($i$) the mass splitting between the resonances should be comparable to 
their widths, \eg, on the order of a few per cent, ($ii$) they share common decay modes  {\it and}, more importantly, ($iii$) the various imaginary entries in their self-energy `matrices' have 
somewhat comparable values. When such a breakdown of the B-W description occurs, the resonances become `coupled' in such a way as to lead to a distortion in their expected combined lineshapes. This 
is caused by the additional interference induced by these off-diagonal terms in the self-energy matrix, which is not captured by the usual B-W prescription. 
Such an interference can be either constructive or destructive in nature depending upon the specifics of the new physics model.  These authors showed that, if 
destructive, it might be possible that the resonance signature can be sufficiently suppressed so as to be entirely missed in first-round collider searches.  
As a specific case in point, in this paper we consider the production at the LHC of the nearly degenerate level-2 KK neutral gauge bosons, 
$Z'(\sim W^0_2)$ and $A'(\sim B^0_2)$, present in the Split-UED{\cite {big}} scenario. As we will see, such states, though at least superficially seeming to satisfy all of the three conditions 
above, do not 
show any significant, non-BW interference effects and certainly none which are sufficient to mask their existence. Before a discussion of the specifics of this Split-UED model, we briefly 
review the essentials of this mixing formalism in the case of two nearby neutral spin-1 resonances.

\section{Formalism and Analysis}

To be explicit, in the discussion presented below we follow the analysis and notations as given in Ref.\cite{Cacciapaglia:2009ic} for the case of two nearby resonances that share 
common (production and ) decay modes. In this $2\times 2$ case, the generalized propagator matrix for these two gauge fields can be written as 
\begin{equation}
i\Delta_{\mu\nu}=-i\Big(g_{\mu\nu}-{{p_\mu p_\nu}\over {p^2}}\Big)~\Delta_s
\,,
\end{equation}
where $i\Delta_s$ is itself the $2\times 2$ propagator matrix for scalars whose elements are given by
\begin{eqnarray}
D_s \Delta_s & =  \left( \begin{array}{cc}
                        p^2-m_2^2+i\Sigma_{22} & -i\Sigma_{12}  \\
                        -i\Sigma_{21} & p^2-m_1^2+i\Sigma_{11}
                  \end{array}\right)\,,
\end{eqnarray}
and where 
\begin{equation}
D_s=(p^2-m_1^2+i\Sigma_{11})(p^2-m_2^2+i\Sigma_{22})+\Sigma_{12}\Sigma_{21}\,.
\end{equation}
The calculation of the quantities $\Sigma_{ij}$ is straightforward from the usual 1-loop vacuum polarization diagrams; in the cases of interest to us, since 
all the couplings of the two neutral gauge fields are real, $\Sigma_{12}=\Sigma_{21}$. These off-diagonal elements will, of course, be zero when the two vectors do not share common 
decay modes. When that happens the matrix is already diagonal and the standard B-W description of the two resonances then goes through as usual. 
Further, we note that these off-diagonal entries need not be positive, \ie, they can quite generally be of either sign. 

Note that in the limit where the masses of all the final state objects in the decays of these 
resonances can be neglected (which will be at least approximately true at LHC energies for the cases of interest) we obtain the momentum scaling 
$\Sigma_{11(22)}\simeq p^2 \Gamma_{1(2)}/m_{1(2)}$ where $\Gamma_i$ are the conventionally calculated on-shell widths of these two resonances. Note that in this 
same limit, $\Sigma_{12}$ will also similarly scale as $\sim p^2$. The $\Sigma_{ij}$ can thus be thought of as generalized running-widths. In the case of 
Split-UED, the final states common to both $W^0_2$ and $B^0_2$ will consist of the usual zero-mode (\ie, SM) fermions. Thus we find that $\Sigma_{12}$ 
can be simply expressed in terms of (weighted) sums, over the SM fermions, of the set of products of the vector and axial-vector couplings of the two 
gauge fields:  $\sum_f (v_1v_2, ~a_1a_2)_f$. It is important to remember that the individual contributions to $\Sigma_{12}$ from any given fermion can have either sign so that the 
{\it overall} sign of $\Sigma_{12}$ is a indeterminate; however, we would a priori expect that the size of this element would be comparable in magnitude to those appearing on the diagonal. 
In the specific numerical analysis below, the $\Sigma_{ij}$ will be calculated including all finite mass effects along with the relevant approximate QCD and QED corrections. 

In order to see how this more complicated propagator structure affects resonance cross sections it is instructive to consider the $s$-channel 
exchange of (in general, $n$) neutral gauge bosons with real couplings between an initial state ($I$) and a final state ($F$) consisting of SM fermions. In 
such a case, in the amplitude in `gauge boson space' can be symbolically written as   
\begin{equation}
{\cal M} = \sum_{ij} I_i \Delta_{ij} F_j\,,
\end{equation}
which leads to
\begin{equation}
{\cal M}{\cal M}^\dagger = \sum_{ijkl} I_i \Delta_{ij} F_j F_k^\dagger \Delta_{kl}^\dagger I_l^\dagger\,,
\end{equation}
or, by simple rearrangement  
\begin{equation}
{\cal M}{\cal M}^\dagger = \sum_{ijkl} (I_iI_l^\dagger)~(F_jF_k^\dagger) ~P_{ijkl}\,,
\end{equation}
where $P_{ijkl}= \Delta_{ij}\Delta_{kl}^\dagger$ and, in the limit of massless fermions, $I_iI_l^\dagger \sim (v_iv_l,~a_ia_l)_{initial}$ and 
$F_jF_k^\dagger  \sim (v_jv_k,~a_ja_k)_{final}$ by taking traces over the gamma matrices and fermion spinors as usual. The denominator of the matrix $P_{ijkl}$ is 
simply $|D_s= det \Delta|^2$ which in this $2\times 2$ case is explicitly given by  
\begin{equation}
|D_s|^2= \Big[(s-m_1^2)(s-m_2^2)+(\Sigma_{12}^2-\Sigma_{11}\Sigma_{22})\Big]^2+\Big[\Sigma_{11}(s-m_2^2)+\Sigma_{22}(s-m_1^2)\Big]^2 \,,
\end{equation}
while the numerator of $P_{ijkl}$ in the $2\times 2$ case effectively contains only 6 independent terms; this follows from hermiticity, the initial symmetries of the propagator matrix itself, 
as well as the reality of the fermion gauge couplings which also simultaneously enforces the cancellation of the (potential) imaginary terms in the sums above.

\section{Split-UED Basics}

The essential details of the properties of Split-UED can be found in Ref.{\cite{big}} which we summarize here. In minimal UED{\cite{Appelquist:2000nn}}{\cite{Cheng:2002iz}}{\cite{Datta:2005zs}},  
SM gauge and matter fields are allowed to propagate freely in a flat, $S^1/Z_2$ orbifolded, 5-D space (of internal radius $R$). These states have the usual sine and cosine type wavefunctions, 
for their various KK-modes, depending upon whether they 
are even or odd under the $Z_2$ symmetry. One of the effects of orbifolding (which is introduced to obtain chiral zero-mode fermions) is to break KK number conservation down to only KK-parity. Thus 
while even mode $n\geq 2$ gauge fields do not couple at tree-level to the zero-mode SM fermions, one-loop radiative corrections can and do induce such small, loop-suppressed couplings. However, unlike 
in minimal UED, in Split-UED the SM fermions are allowed to have bulk mass terms, $\mu_i \sim R^{-1}$ which lead to a distortions in both the fermion KK spectrum and the associated wavefunctions. 
In particular, zero-mode fermions now have wavefunctions which either peak at the center ($y=0$) or at the boundaries ($y=\pm L=\pm \pi R/2$) of the orbifold depending upon the the sign of 
$\mu$. One effect of this, particularly relevant for our discussion here, is to allow for a direct tree-level coupling of the even $n=2m$ KK gauge fields to the SM zero-mode fermions. 

For a fermion 
with a bulk mass $x=\mu L$ the coupling to the $(2m)^{th}$-mode gauge field for $m>0$ in units of the corresponding SM gauge coupling is given by the function{\cite{Rizzo:2001sd}}      
\begin{equation}
{\cal F}_{002m}= {{x^2~[1-(-1)^m e^{2x}]~[1-\coth (x)]}\over {\sqrt 2 ~[x^2+(m\pi/2)^2]}}\,,
\end{equation}
which vanishes as $x\to 0$ (corresponding to the usual tree-level minimal UED limit) and goes to $(-1)^m\sqrt 2$ as $x\to \infty$ as the fermion becomes highly localized at the origin. For simplicity 
and purposes of demonstration in the discussion below we will taken a common value for $x\sim 1$ for all of the SM fermions. Precision electroweak (EWK)
 constraints indicate that the region of low $x \lsim 1$ 
is now somewhat preferred{\cite{Flacke:2011nb}}{\cite{Huang:2012kz}}. 
For such typical parameter values, the width to mass ratios of both the $W^0_2$ and $B^0_2$ are of order a few percent while their 
loop-induced mass splitting{\cite{Cheng:2002iz}} is found to  be comparable, on the order of $\sim 6\%$. As they certainly share the various SM fermions as common decay modes these two gauge 
KK fields apparently 
seem to meet all of the criteria $(i)-(iii)$ above for states which may have important non-BW interference. We now turn to a numerical study of these effects at the LHC.

\section{Numerical Investigation and Results}

Searches for new neutral gauge bosons at the 7 TeV LHC in the Drell-Yan channel have been performed by both the ATLAS and CMS collaborations{\cite {ATLAS,CMS,moriond}} but with  
only null results so far. Preliminary CMS results at 8 TeV also show no hint or a signal{\cite {CMS2}} with $\sim 5$ fb$^{-1}$ of luminosity. These searches have become fairly powerful, already 
excluding a $Z'$ with SM-like couplings below a mass 
of $\simeq 2.3-2.6$ TeV with an integrated luminosity of $\simeq 5$ fb$^{-1}$. Of course these limits would degrade for somewhat weaker couplings and/or smaller branching fractions into the dilepton 
final state. Thus the first question we should address is whether or not width mixing effects could be hiding Split-UED in current data and then to address the issue of whether or not future data 
at higher energies and luminosities could also hide such a signal.

In our case of interest, once the values of $(R,x,\Lambda)$,  
with $\Lambda R =20$ being the cutoff scale, are chosen all of couplings and other properties of the new gauge bosons are completely specified. Recall that due to EWK breaking and radiative 
corrections the weak eigenstates $W_2^0$ and $B_2^0$ are only {\it approxomate} mass eigenstates{\cite{Cheng:2002iz}} and experience a relatively small amount of mass mixing via a calculable angle, 
$\phi$ which approximately scales as $\sim (vR)^2$, with $v$ being the usual Higgs vev, in both the neutral and charged gauge boson sectors.  
This result follows immediately from the calculations of the various KK gauge boson masses (including the contributions of important loop corrections) which are 
clearly left unaltered in the current Split-UED scenario{\cite{Cheng:2002iz}}. {\footnote {Ref.\cite{Flacke:2008ne} has more generally shown that such a relation is to be expected in many other extensions 
of UED.}}  We make use of the exact expressions for all of the various masses and mixing angles in the analysis that follows. 
Since we must fully account for these small but important mixing effects in our analysis we we will instead refer to the physical eigenstates as $Z' \simeq W_2^0$ and $A'\simeq B_2^0$ to avoid 
any potential confusion. In this basis, the $Z'$ couples to the SM fermions as 
\begin{equation}
F~(g/c_w)~[(cc_w+ss_w)~T_3-ss_w~Q]\,,
\end{equation}
while $A'$ couples to these same fermions instead as 
\begin{equation}
F~(g/c_w)~[(sc_w-cs_w)~T_3+cs_w~Q]\,,
\end{equation}
where $F={\cal F}_{002}\sim 1$, $T_3$ is the usual isospin generator, $Q$ is the electric charge, $s_w(c_w)= \sin \theta_w(\cos \theta_w)$ in terms of the usual weak mixing angle and 
$c(s)= \cos \phi (\sin \phi)$. Complete expressions for the relevant mass matrices from Ref.{\cite{Cheng:2002iz}} are used in our numerical analysis below. 
Note that when $\phi \to \theta_w$ and $F=1$, as will occur for zero-mode gauge fields, we recover the usual SM gauge couplings for the $Z$ and $A$. However 
here we find that $s\sim (0.5~RM_Z)^2 <<1$ for the $n=2$ modes. In order to perform our numerical calculations for the 
LHC we will employ CTEQ6.6 parton densities{\cite{Pumplin:2002vw}} and approximate NNLO K-factors{\cite{Melnikov:2006kv}}. As we noted above, the various SM fermion partial width calculations 
will include finite mass effects as well as approximate NLO QCD and LO QED radiative corrections.

\begin{figure}[htbp]
\centerline{
\includegraphics[width=8.5cm,angle=90]{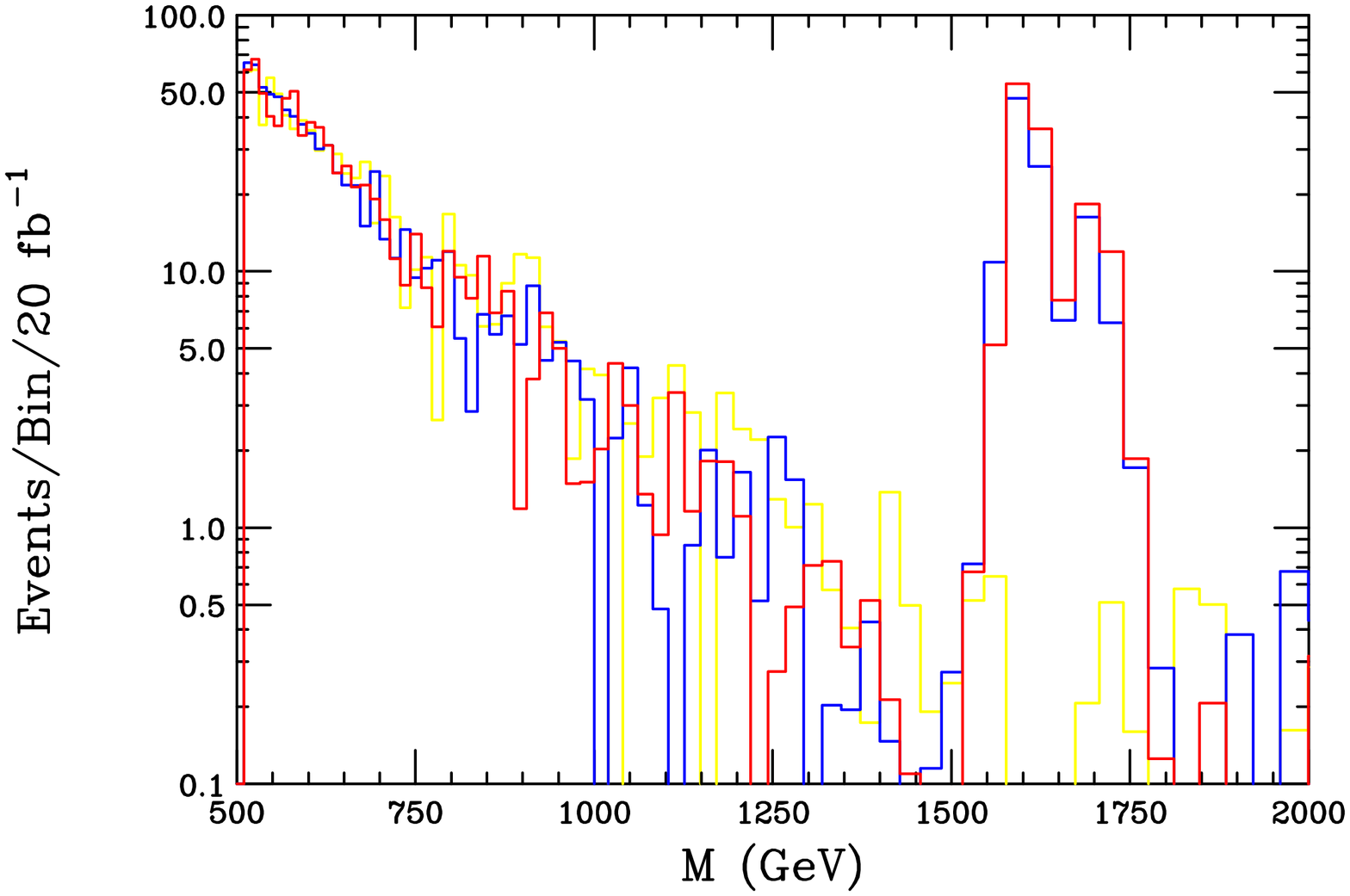}}
\vspace*{0.5cm}
\centerline{
\includegraphics[width=8.5cm,angle=90]{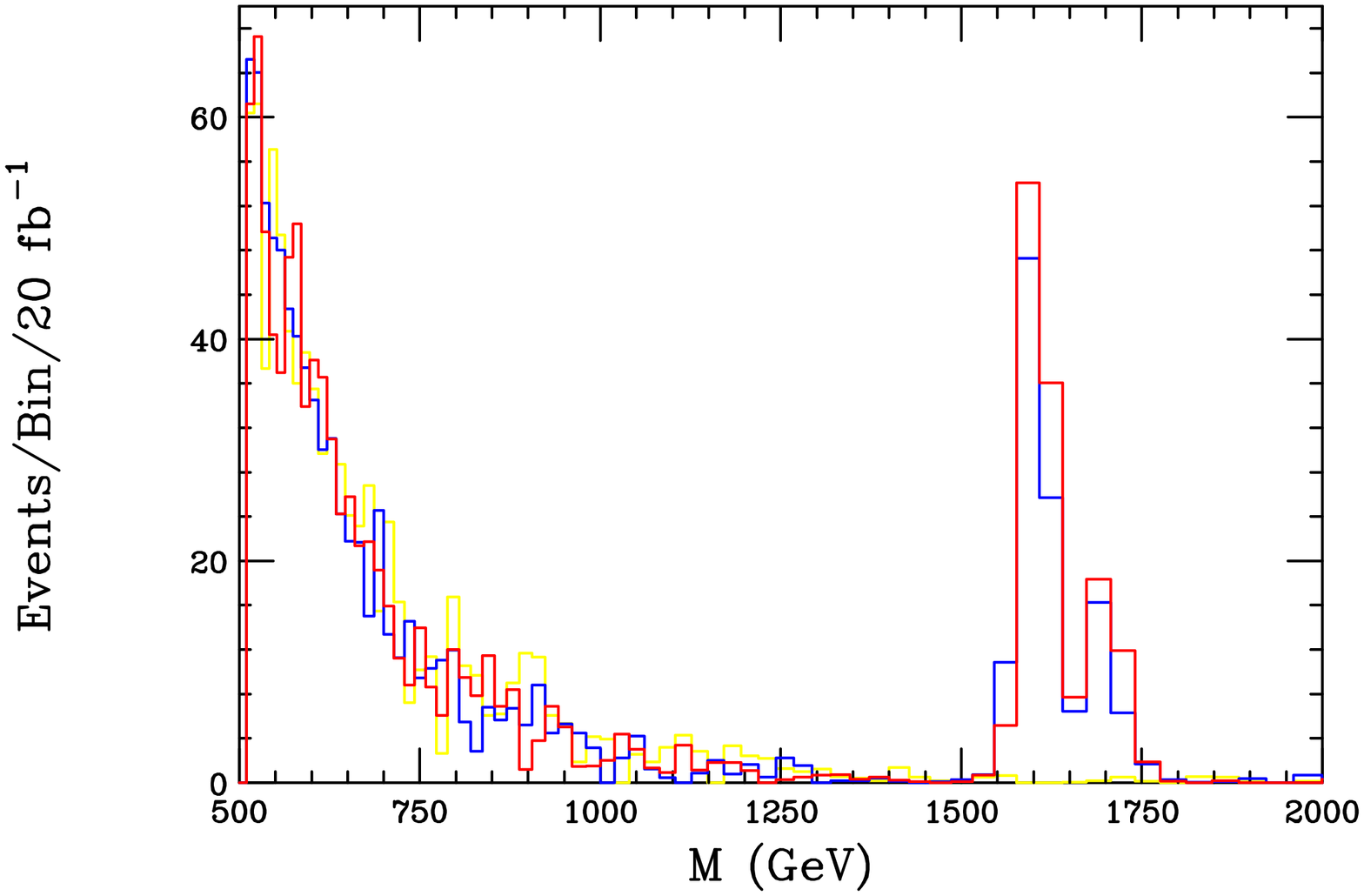}}
\vspace*{0.2cm}
\caption{Results on both log (top) or linear (bottom) scales for $\sqrt s=8$ TeV and an integrated luminosity of 20 fb$^{-1}$ assuming $R^{-1}=800$ GeV and $x=1$. The yellow histogram is the SM background 
from conventional Drell-Yan production while the blue (red) histogram shows the expectations for Split-UED without (with) the effects of width mixing included. The results have been smeared by the 
$\sim 1\%$ mass resolution of the ATLAS detector.}
\label{fig1}
\end{figure}

To be specific for purposes of demonstration, let us focus on the important process $pp \to e^+e^- +X$ assuming $R^{-1}=800$ GeV and $x=1$ at the $\sqrt s=8$ TeV LHC with an integrated luminosity of 
20 fb$^{-1}$. Under such conditions we would anticipate that the double resonance signal structure would be quite obvious and this is indeed the case as is shown in Fig.~\ref{fig1}. The result shown 
here reflects our current ($\sqrt s=8$ TeV with a luminosity of 5 fb$^{-1}$) knowledge for $Z'$-like searches. Furthermore, at 
the level of the statistical fluctuations we see here the rather {\it unanticipated} result that there is very little if any observable difference between the conventional Split-UED signal and the one in which 
the width mixing effects are included. We thus conclude that such width mixing effects would not be able to hide Split-UED signatures in the current LHC data sample.  

\begin{figure}[htbp]
\centerline{
\includegraphics[width=8.5cm,angle=90]{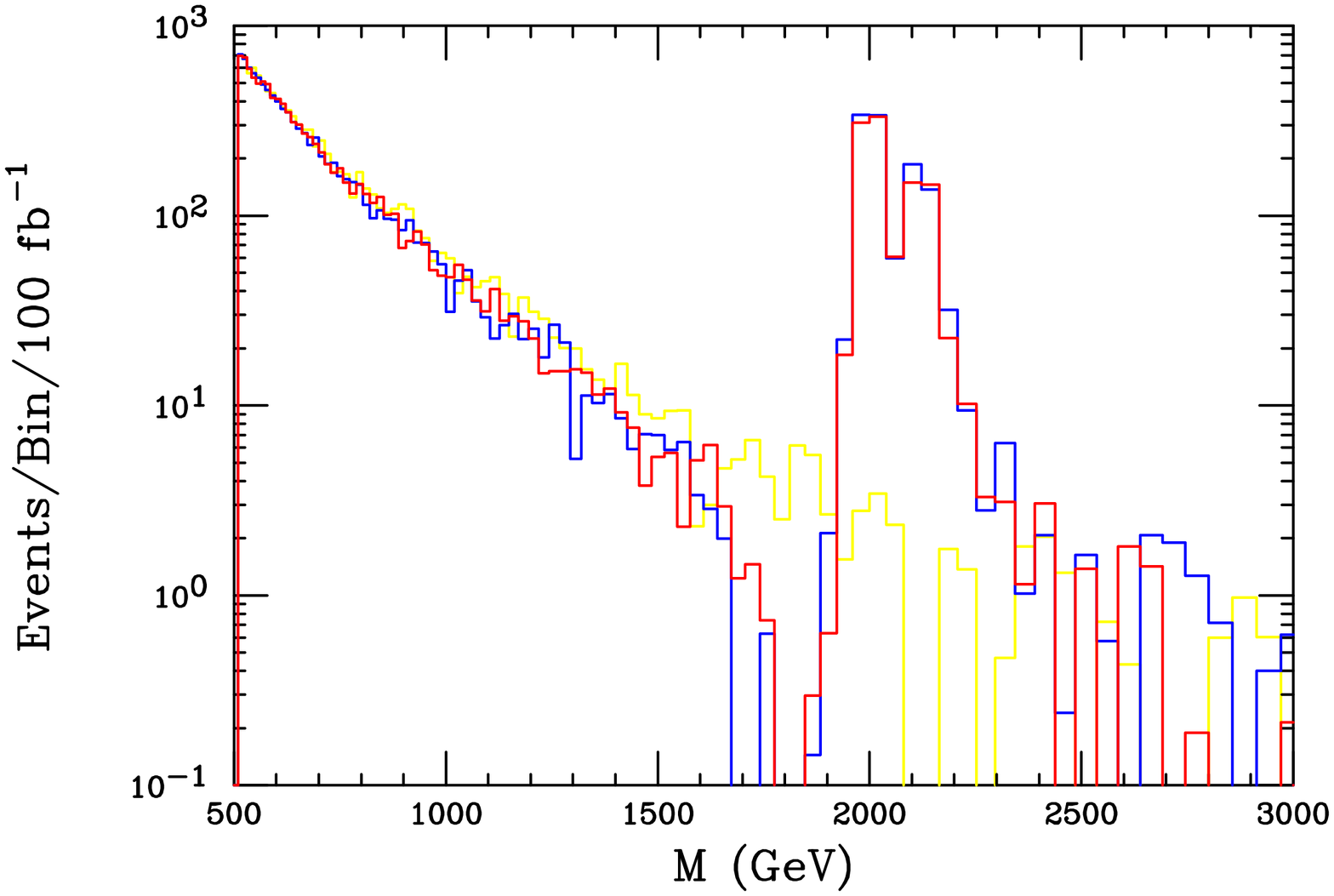}}
\vspace*{0.5cm}
\centerline{
\includegraphics[width=8.5cm,angle=90]{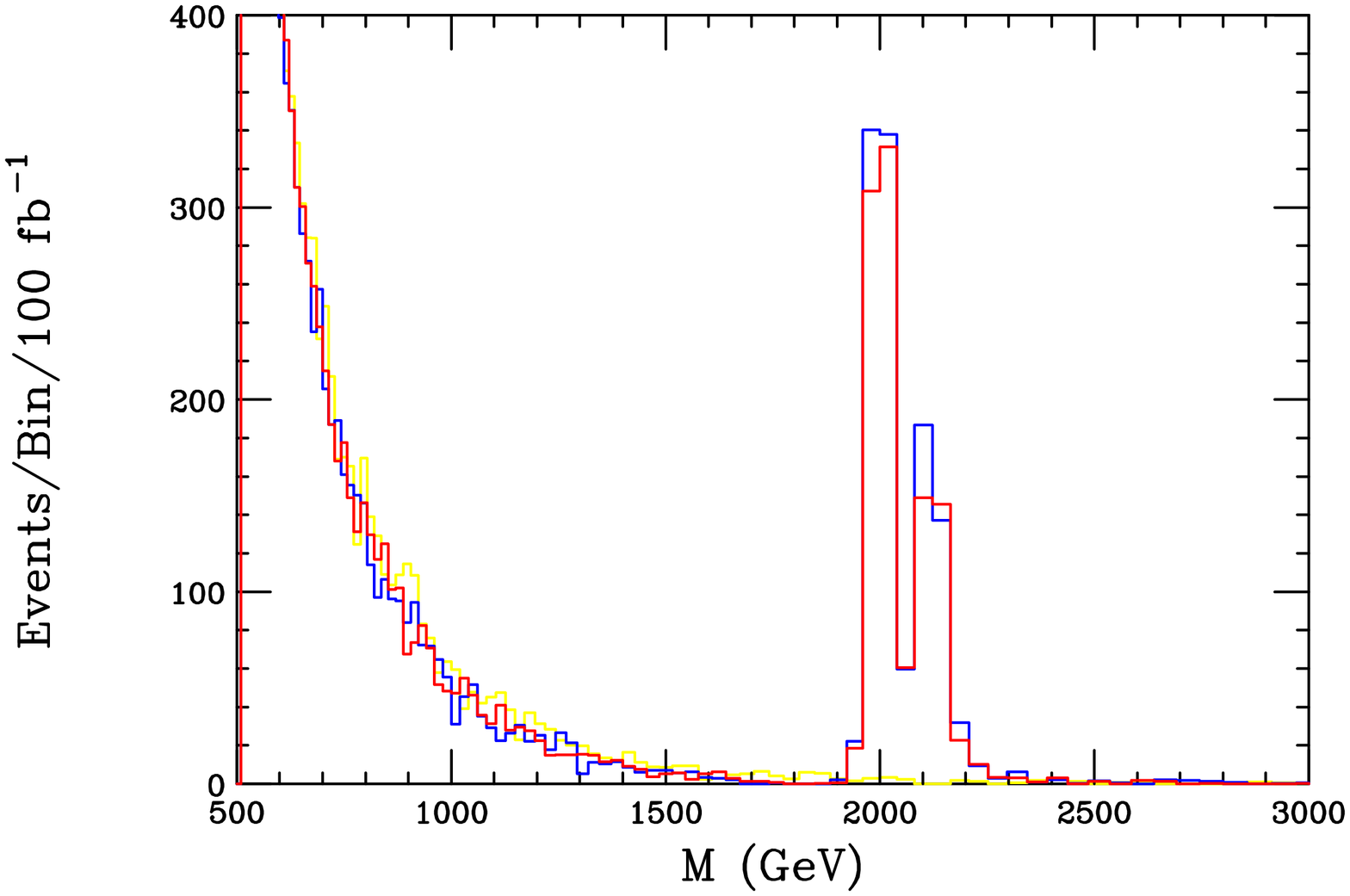}}
\vspace*{0.2cm}
\caption{Same as the previous figure but now for $\sqrt s=13$ TeV and an integrated luminosity of 100 fb$^{-1}$ assuming $R^{-1}=1$ TeV and $x=1$.}
\label{fig2}
\end{figure}

To see whether or not this is just an effect of limited statistics and to see what the LHC may be able to do in the future in this regard, we increase $R^{-1}$ to 1 TeV, $\sqrt s$ to 13 TeV and the luminosity to 
100 fb$^{-1}$ keeping $x=1$ and show the corresponding result in Fig.~\ref{fig2}. Again we see that the double resonance structure is clearly visible with or without the inclusion of width mixing 
effects (as is the usual strong destructive interference below the peaks signaling KK gauge boson production). Clearly this is no longer an issue of statistics but the actual absence of  
any visible width mixing effects contrary to our expectations. How can such an outcome be realized if we seemingly satisfied all of the necessary conditions above? Is it a numerical accident or something deeper?


Actually what one finds, under closer examination, is that while all the above criteria for a strong width mixing effect would seem to be satisfied, in reality they are not.  In the specific 
case at hand, even though the two gauge KK states are quite close in mass and share many common decay modes via the SM fermions, the size of the {\it total} off-diagonal decay width remains quite small 
in comparison to the diagonal ones so that the 
two resonances remain effectively decoupled into ordinary B-W states. That this can happen numerically is the result of the fact that the contributions from the various SM fermions to this off-diagonal 
width can appear with either sign. When these various contributions are then summed in the present case, however, the total is found to be quite small, in fact, near zero. Why is this? 

To understand what is happening consider for simplicity of discussion the value of $\Sigma_{12}$ in the limit where all the SM fermion masses can be neglected relative to those of $Z'$ and $A'$ (this is 
an excellent approximation) and where the leading QCD and QED corrections to the various partial widths can also be neglected. (Of course these approximations are introduced only 
for the ease of this discussion and are not used in obtaining any of the numerical results that are shown here.) 
In such a limit, we can express the value of $\Sigma_{12}$ in terms of a trace over products of the SM generators 
$T_3$ and $Y/2$ using the coupling expressions for $Z'$ and $A'$ above and by remembering that in the SM $Q=T_3+Y/2$. Explicitly, one finds that 
\begin{equation}
\Sigma_{12}\sim  Tr_f ~\big[sc(c_w^2~T_3^2-s_w^2~(Y/2)^2)+s_wc_w~(c^2-s^2)~(T_3 Y/2) \big]\,,
\end{equation}
where the trace is taken over the various SM fermions. Although $Tr_f (T_3)^2=2$ and $Tr_f (Y/2)^2=10/3$ for each generation and are thus of order unity, the first term in the square bracket is highly 
suppressed due to the very small value of the angle $\phi$, which here is of order $s\sim (0.5~M_ZR)^2 <<1$. {\footnote {Interestingly, at the conventional GUT scale the corresponding term in 
parentheses would also vanish due to the GUT normalization condition on the SM 
generators. At low scales this means that the numerical value of this quantity is somewhat smaller than one might have naively guessed.}} 
The second term has no such suppression factor, however, since here the coefficient $c^2-s^2 \sim 1$ appears. But in this case we remember that $Tr_f~ T_3 Y/2=0$ since all members of any weak isospin multiplet 
will necessarily have the same value of the weak hypercharge so that this term must automatically vanish (as for any isomultiplet $Tr T_3=0$). Thus as long as $\phi$ is very small we can never generate a 
sufficiently large off-diagonal element in the width matrix to produce a statistically significant non-B-W effect. This remains true even if new fields are added to those already occuring in the SM or if one 
modifies our specific choices of the basic Split-UED input parameters.  There are no significant width mixing effects in the case of gauge boson KK production in Split-UED as long as the fermionic couplings 
are given as above and this result is not significantly modified by the small modifications due to finite fermion masses or QED/QCD corrections. The reason for this is that the leading order QCD corrections 
are flavor-independent and that the largest fermion mass effect is $\sim (m_t/M_{Z'})^2 \sim s$, \ie, the same size as the small mixing terms discussed above. Further, note that since the width mixing takes 
place solely within the gauge boson propagators and the masses of the incoming quarks are effectively zero (so that the longitudinal part of the gauge propagators can be dropped) we will observe the same 
(lack of) width mixing effect for all other possible final states such as $\tau^+\tau^-, b\bar b$ or $t\bar t$.

\section{Discussion and Conclusion}

In this paper we have examined the possible influence of non-B-W width mixing effects on the production of a nearly degenerate pair of level-2 KK gauge bosons, $Z',A'$, at the LHC within the framework of 
Split-UED. Although this model apparently seems to satisfy all of the criterion necessary for such effects to be numerically significant, surprisingly, on closer inspection it does not. One finds that the total 
contribution to the width mixing parameter is very highly suppressed and even vanishes in the limit when small mass mixing between the $Z'$ and $A'$ states goes to zero.  The main suppression of this width 
mixing effect is due to the orthogonal nature of the $T_3$ and $Y/2$ generators in the SM, \ie, the fact that $Tr_f ~T_3Y/2=0$ identically for any weak isospin representation. Unfortunately, although the 
$Z'$ and $A'$ are nearly degenerate and share common decay modes, when summed over the off-diagonal partial widths essentially add up to (almost) zero, thus producing an insignificant width mixing effect. 
Hopefully the LHC will provide us some other laboratory to study effects of this kind.

\noindent{\Large\bf Acknowledgments}

The author would like to thank J.L. Hewett for discussions related to this work.

%
\def\MPL #1 #2 #3 {Mod. Phys. Lett. {\bf#1},\ #2 (#3)}
\def\NPB #1 #2 #3 {Nucl. Phys. {\bf#1},\ #2 (#3)}
\def\PLB #1 #2 #3 {Phys. Lett. {\bf#1},\ #2 (#3)}
\def\PR #1 #2 #3 {Phys. Rep. {\bf#1},\ #2 (#3)}
\def\PRD #1 #2 #3 {Phys. Rev. {\bf#1},\ #2 (#3)}
\def\PRL #1 #2 #3 {Phys. Rev. Lett. {\bf#1},\ #2 (#3)}
\def\RMP #1 #2 #3 {Rev. Mod. Phys. {\bf#1},\ #2 (#3)}
\def\NIM #1 #2 #3 {Nuc. Inst. Meth. {\bf#1},\ #2 (#3)}
\def\ZPC #1 #2 #3 {Z. Phys. {\bf#1},\ #2 (#3)}
\def\EJPC #1 #2 #3 {E. Phys. J. {\bf#1},\ #2 (#3)}
\def\IJMP #1 #2 #3 {Int. J. Mod. Phys. {\bf#1},\ #2 (#3)}
\def\JHEP #1 #2 #3 {J. High En. Phys. {\bf#1},\ #2 (#3)}

\end{document}